\documentclass[twocolumn]{aastex631}

\pdfoutput=1
\usepackage{graphicx}                      
\DeclareGraphicsExtensions{.pdf,.png,.jpg} 

\begin{document}

\title{ELPIS: Accelerated metal and dust enrichment in a proto-cluster core at $z\approx8$}
\author{Hideki Umehata}
\affiliation{Institute for Advanced Research, Nagoya University, Furocho, Chikusa, Nagoya 464-8602, Japan}
\affiliation{Department of Physics, Graduate School of Science, Nagoya University, Furocho, Chikusa, Nagoya 464-8602, Japan}

\author{Yoichi Tamura}
\affiliation{Department of Physics, Graduate School of Science, Nagoya University, Furocho, Chikusa, Nagoya 464-8602, Japan}

\author{Yoshinobu Fudamoto}
\affiliation{Center for Frontier Science, Chiba University, 1-33 Yayoi-cho, Inage-ku, Chiba 263-8522, Japan}

\author{Yurina Nakazato}
\affiliation{Department of Physics, The University of Tokyo, 7-3-1 Hongo, Bunkyo, Tokyo 113-0033, Japan}

\author{Daniel Ceverino}
\affiliation{Universidad Autonoma de Madrid, Ciudad Universitaria de Cantoblanco, E-28049 Madrid, Spain}
\affiliation{CIAFF, Facultad de Ciencias, Universidad Autonoma de Madrid, E-28049 Madrid, Spain}

\author{Naoki Yoshida}
\affiliation{Department of Physics, The University of Tokyo, 7-3-1 Hongo, Bunkyo, Tokyo 113-0033, Japan}
\affiliation{Kavli Institute for the Physics and Mathematics of the Universe (WPI), UT Institute for Advanced Study,
The University of Tokyo, Kashiwa, Chiba 277-8583, Japan}
\affiliation{Research Center for the Early Universe, School of Science, The University of Tokyo, 7-3-1 Hongo, Bunkyo,
Tokyo 113-0033, Japan}

\author{Akio K. Inoue}
\affiliation{Waseda Research Institute for Science and Engineering, Faculty of Science and Engineering, Waseda University,
3-4-1 Okubo, Shinjuku, Tokyo 169-8555, Japan}
\affiliation{Department of Pure and Applied Physics, School of Advanced Science and Engineering, Faculty of Science
and Engineering, Waseda University, 3-4-1 Okubo, Shinjuku, Tokyo 169-8555, Japan}

\author{Ryota Ikeda}
\affiliation{Department of Astronomy, School of Science, SOKENDAI (The Graduate University for Advanced Studies), 2-21-1 Osawa, Mitaka, Tokyo 181-8588, Japan}
\affiliation{National Astronomical Observatory of Japan, 2-21-1 Osawa, Mitaka, Tokyo 181-8588, Japan}

\author{Yuma Sugahara}
\affiliation{Waseda Research Institute for Science and Engineering, Faculty of Science and Engineering, Waseda University,
3-4-1 Okubo, Shinjuku, Tokyo 169-8555, Japan}
\affiliation{Department of Pure and Applied Physics, School of Advanced Science and Engineering, Faculty of Science
and Engineering, Waseda University, 3-4-1 Okubo, Shinjuku, Tokyo 169-8555, Japan}

\author{Shutaro Inui}
\affiliation{Department of Physics, Graduate School of Science, Nagoya University, Furocho, Chikusa, Nagoya 464-8602, Japan}

\author{Santiago Arribas}
\affiliation{Centro de Astrobiolog\'{\i}a (CAB), CSIC-INTA, Ctra. de Ajalvir km 4, Torrej´on de Ardoz, E-28850, Madrid, Spain}

\author{Tom Bakx}
\affiliation{Department of Space, Earth, \& Environment, Chalmers University of Technology, Chalmersplatsen,
SE-4 412 96 Gothenburg, Sweden}

\author{Masato Hagimoto}
\affiliation{Department of Physics, Graduate School of Science, Nagoya University, Furocho, Chikusa, Nagoya 464-8602, Japan}

\author{Takuya Hashimoto}
\affiliation{Division of Physics, Faculty of Pure and Applied Sciences, University of Tsukuba, Tsukuba, Ibaraki 305-8571, Japan}

\author{Luis Colina}
\affiliation{Centro de Astrobiolog\'{\i}a (CAB), CSIC-INTA, Ctra. de Ajalvir km 4, Torrej´on de Ardoz, E-28850, Madrid, Spain}

\author{Yi W. Ren}
\affiliation{Department of Pure and Applied Physics, School of Advanced Science and Engineering, Faculty of Science
and Engineering, Waseda University, 3-4-1 Okubo, Shinjuku, Tokyo 169-8555, Japan}

\author{Wataru Osone}
\affiliation{Division of Physics, Faculty of Pure and Applied Sciences, University of Tsukuba, Tsukuba, Ibaraki 305-8571, Japan}

\author{Alejandro Crespo G\'omez}
\affiliation{Space Telescope Science Institute (STScI), 3700 San martin Drive, Baltimore, MD 21218, USA }

\author{Ken Mawatari}
\affiliation{Waseda Research Institute for Science and Engineering, Faculty of Science and Engineering, Waseda University,
3-4-1 Okubo, Shinjuku, Tokyo 169-8555, Japan}
\affiliation{Department of Pure and Applied Physics, School of Advanced Science and Engineering, Faculty of Science
and Engineering, Waseda University, 3-4-1 Okubo, Shinjuku, Tokyo 169-8555, Japan}

\author{Javier {\'A}lvarez-M{\'a}rquez}
\affiliation{Centro de Astrobiolog\'{\i}a (CAB), CSIC-INTA, Ctra. de Ajalvir km 4, Torrej´on de Ardoz, E-28850, Madrid, Spain}



\begin{abstract}
We present a study of the metal, dust, and molecular gas content in galaxies within the A2744-z7p9OD proto-cluster at $z \approx 7.88$. We focus on two galaxy groups—the {\it Quintet} and the {\it Chain}—which are covered by the “ELPIS” survey (The Emission-Line Protocluster Imaging Survey of the furthest overdensity
beyond Pandora's Cluster Abell 2744). [C\,\textsc{ii}] 158\,$\mu$m emission is detected in five galaxies, revealing molecular gas reservoirs with ${\rm log}(M_{\rm gas}/M_\odot) \sim 9.0$--9.6, while dust continuum at the observed frame of 1.26\,mm is detected in three galaxies, yielding dust masses of ${\rm log}(M_{\rm dust}/M_\odot) \sim 6.0$--6.4, assuming a dust temperature of $T_{\rm dust} = 45^{+15}_{-15}$\,K. 
The derived properties—including stellar-to-dust mass ratios of $\log (M_{\rm dust}/M_\star) \sim -3$ to $-2$ at $\log (M_\star/M_\odot) \approx 9$, and dust-to-gas mass ratios of $\log (M_{\rm dust}/M_{\rm gas}) \sim -4$ to $-3$ at $12+\log ({\rm O/H}) \approx 8$—place these galaxies in an intermediate regime: higher than the very low ratios expected from supernova-driven dust production, but still below the levels attained once efficient grain growth dominates. These values indicate a transition phase of dust mass assembly, likely reflecting the onset of grain growth via metal accretion under accelerated evolution in the proto-cluster core.
\end{abstract}




\section{Introduction} \label{sec:intro}

\begin{figure*}
  \centering
  \includegraphics[width=1.8\columnwidth]{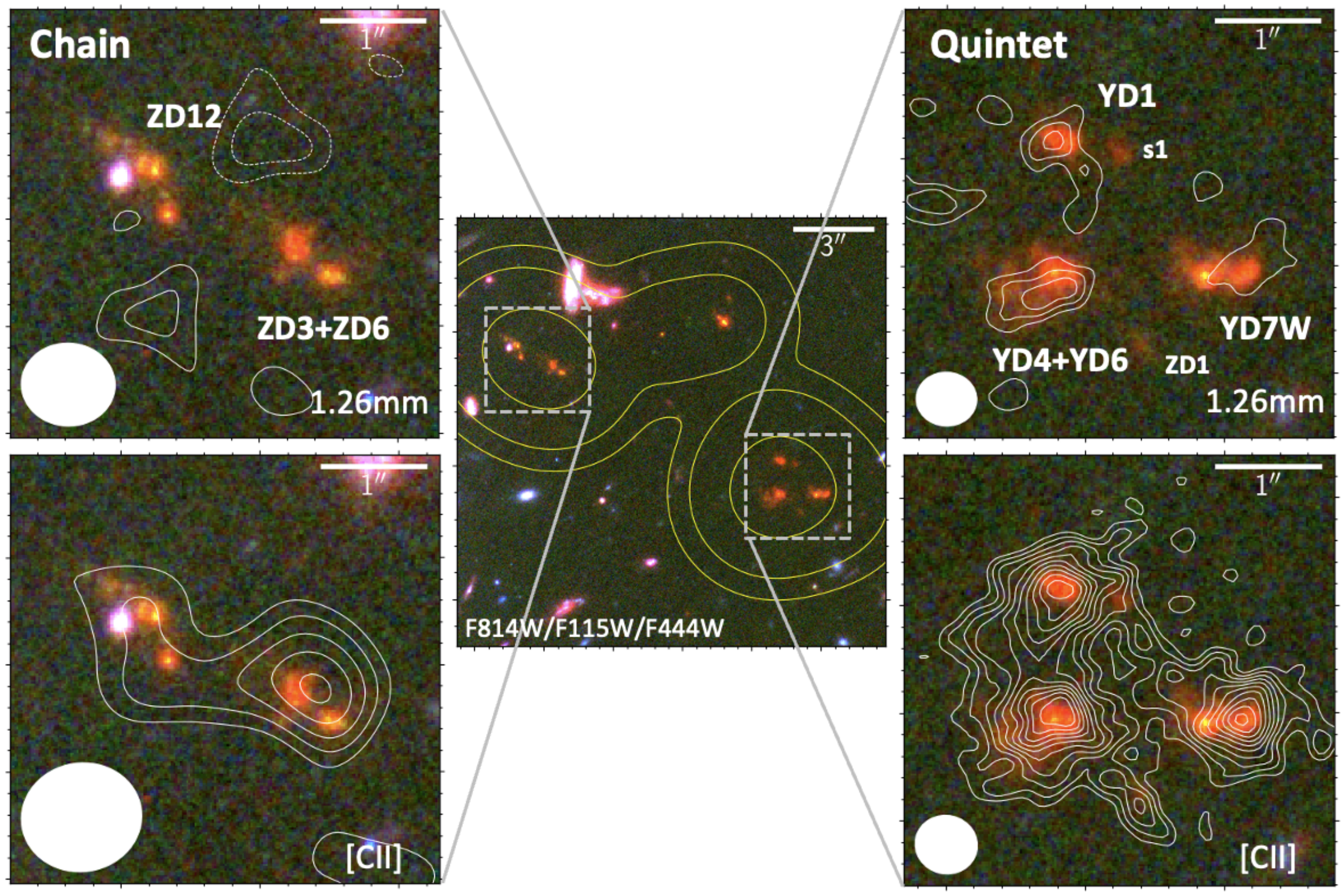}
\caption{
The middle panel shows an HST+JWST color composite image (F814W/F115W/F444W) of a portion of the A2744-z7p9OD proto-cluster, with F814W assigned to blue, F115W to green, and F444W to red, at $z \approx 7.88$. Yellow contours indicate the stellar-mass-weighted density map constructed from spectroscopically confirmed member galaxies, corresponding to 55\%, 75\%, and 95\% of the peak mass density \citep{2025arXiv250706284W}. The left and right panels show enlarged views of the {\it Chain} and {\it Quintet} groups, respectively, with overlaid 1.26\,mm dust continuum (top) and [C\,\textsc{ii}] line (bottom) white contours starting at $\pm2\sigma$ and increasing in steps of $1\sigma$. Naturally weighted maps are used for the {\it Quintet}, while tapered images are shown for the {\it Chain}.} [C\,\textsc{ii}] emission is detected from five galaxies across both regions, while dust continuum is detected only in the {\it Quintet}.
\label{fig:images}
\end{figure*}

Dust production in the early universe is a key process that informs our understanding of the chemical evolution of galaxies and the enrichment of the interstellar and circumgalactic medium (ISM and CGM). Recent observations with the Atacama Large Millimeter/submillimeter Array (ALMA) have revealed that massive dust reservoirs are already in place in galaxies at $z \approx 7$--8 \citep[e.g.,][]{2015Natur.519..327W,2018Natur.553...51M,2019PASJ...71...71H,2019ApJ...874...27T,2023ApJ...952....9T,2020MNRAS.493.4294B,2021ApJ...921...97J,2023MNRAS.518.6142A,2024MNRAS.533.3098A,2025arXiv250110508A,2022MNRAS.515.1751W}. These findings imply that early galaxies experience rapid metal enrichment, accompanied by efficient dust formation on short timescales.

Dust forms through the condensation of metals expelled by evolved stars. In the local universe, the primary sources of dust are the stellar winds of asymptotic giant branch (AGB) stars and the ejecta from Type II supernovae (SNe II). However, given the relatively long evolutionary timescales of AGB stars, they are thought to play a minor role in global dust production at $z \gtrsim 6$ \citep[e.g.,][]{2015MNRAS.451L..70M,2022MNRAS.512..989D}.
At such a high redshift, dust growth via metal accretion in the ISM is thought to become a key pathway, potentially rivaling supernovae as the dominant mechanism of dust mass buildup. This process requires an initial dust and metal seed population—most likely supplied by early SNe—after which gas-phase metals can accrete onto existing grains, thereby increasing the total dust mass \citep[e.g.,][]{2012MNRAS.422.1263H,2013EP&S...65..213A}.
To fully understand this process, it is critical to probe the multi-phase ISM in early galaxies by simultaneously measuring gas-phase metallicity, dust, and molecular gas. This enables us to trace the evolution of baryonic matter through quantities such as the dust-to-stellar mass ratio and dust-to-gas mass ratio in the early universe.

\citet{2025arXiv250110508A} recently compiled measurements of these quantities for REBELS galaxies at $z\sim7$ and argued that the presence of massive dust reservoirs in these systems requires both rapid metal enrichment by supernovae and efficient dust growth in the ISM. In this Letter, we extend such efforts to an overdensity at $z\sim8$, targeting a regime of lower stellar mass and metallicity at even earlier cosmic time. Given the strong dependence of dust production on environment observed at $z \sim 3$--5—particularly within proto-cluster cores \citep[e.g.,][]{2015ApJ...815L...8U,2019Sci...366...97U,2018ApJ...856...72O,2018Natur.556..469M}
—we also investigate the role of large-scale structure in shaping early dust and metal enrichment in the very early cosmic time.
Throughout this work, we adopt a standard $\Lambda$CDM cosmology with $H_0 = 70$\,km\,s$^{-1}$\,Mpc$^{-1}$, $\Omega_{\rm m} = 0.30$, and $\Omega_\Lambda = 0.70$.

\section{Observation and Reduction} \label{sec:obs}

Our target is the A2744-z7p9OD protocluster at $z = 7.88$, which hosts over ten spectroscopically confirmed member galaxies identified with JWST \citep{2023ApJ...947L..24M,2025ApJ...985...83M,2023ApJ...955L...2H,2024A&A...691A..19V,2025arXiv250706284W}. Within this structure, we focus on two galaxy groups—{\it Chain} and {\it Quintet}—located within the protocluster core (Figure~\ref{fig:images}).

We used ALMA Band~6 to observe the redshifted [C\,\textsc{ii}] 158\,$\mu$m line and underlying dust continuum in the A2744-z7p9OD protocluster. Observations were obtained from three ALMA programs.
Program \#2023.1.01362.S (PI: Y.~Tamura), part of the ``ELPIS'' survey (The Emission-Line Protocluster Imaging Survey of the furthest overdensity
beyond Pandora's Cluster Abell 2744; Y.~Tamura et al. in prep.), was conducted between 2024 January and April using 34--46 antennas under PWV conditions of 1.7--3.3\,mm. The survey employed seven pointings, fully covering both the {\it Chain} and {\it Quintet} groups. The correlator was configured with two spectral windows (213.53--216.39\,GHz and 226.06--229.79\,GHz); one SPW was set to 4-bit mode to target [C\,\textsc{ii}], while others used 2-bit FDM. The total on-source time for the mosaic was 15.6\,hr.
Program \#2023.1.00193.S (PI: Y.~Fudamoto) was carried out between 2024 June 2--12 with 41--47 antennas (PWV = 1.3--2.0\,mm). The pointing was centered at ($\alpha$, $\delta$) = (00$^\mathrm{h}$14$^\mathrm{m}$24\fs9, $-30^\circ$22$'$56\farcs1), corresponding to the region around the {\it Quintet}. The correlator covered 211.43--215.18\,GHz and 226.31--230.06\,GHz using four SPWs in FDM mode. The on-source time was 11.4\,hr.
Program \#2018.1.01332.S (PI: N.~Laporte) provided ancillary continuum data. Observations were conducted in 2019 May with 41--44 antennas (PWV = 0.7--1.3\,mm), using SPWs centered at 245.39--247.2\,GHz and 259.20--261.12\,GHz. The total on-source time was 5.2\,hr.

Data reduction was carried out using CASA v6.5.4 \citep{2007ASPC..376..127M,2022PASP..134k4501C}. The image cube was produced with the \texttt{tclean} task using natural weighting and auto-masking down to a $2\sigma$ threshold. The resulting synthesized beam is $0\farcs58 \times 0\farcs54$ (PA = $77^\circ$), with a typical rms of 61\,$\mu$Jy\,beam$^{-1}$ per 40\,km\,s$^{-1}$ channel at $\sim$214\,GHz.
A tapered cube was also generated using \texttt{uvtaper} = $0\farcs5$, yielding a beam of $1\farcs12 \times 1\farcs01$ (PA = $-81^\circ$) and a typical noise level of 82\,$\mu$Jy\,beam$^{-1}$ per 40\,km\,s$^{-1}$.
We tested continuum subtraction with \texttt{imcontsub} and obtained results consistent with those without subtraction. To avoid introducing additional systematics, we therefore did not apply continuum subtraction in the final analysis.
Continuum images were derived from dedicated measurement sets with the [C\,\textsc{ii}] line channels masked. We used multi-frequency synthesis in \texttt{tclean} with a $2\sigma$ deconvolution threshold, producing two 237.22\,GHz (1.26\,mm) continuum maps: (1) untapered, with a beam of $0\farcs56 \times 0\farcs50$ (PA = $87^\circ$) and rms = 3.5\,$\mu$Jy\,beam$^{-1}$; and (2) tapered (\texttt{uvtaper} = $0\farcs4$), with a beam of $0\farcs88 \times 0\farcs77$ (PA = $87^\circ$) and rms = 3.8\,$\mu$Jy\,beam$^{-1}$.

\section{Results} \label{sec:results}

[C\,\textsc{ii}] emission was searched for within the data cubes, extracting spectra for the regions, yielding detections in both galaxy groups. For the {\it Quintet}, the velocity-integrated flux map over $\Delta v = 360$\,km\,s$^{-1}$ is shown in the lower-right panel of Figure~\ref{fig:images}. Three [C\,\textsc{ii}] peaks are identified, spatially coincident with known proto-cluster members YD1, YD4+YD6, and YD7W. The emission also shows spatial extension that bridges these massive systems like the SSA22-LAB1 at $z=3.09$ \citep{2021ApJ...918...69U}, encompassing lower-mass members such as s1 and ZD1. Detailed analyses of these extended components will be presented in forthcoming papers (Y.~Fudamoto et al., in prep.; Y.~Tamura et al., in prep.). In this Letter, we focus on the three main systems. Since YD4 and YD6 are not spatially separable in [C\,\textsc{ii}] emission, we treat them as a single source (YD4+YD6). Integrated line fluxes were measured for each source using the \texttt{imfit} task in CASA. For the {\it Chain}, we used the tapered cube to enhance the signal-to-noise ratio (S/N) of the relatively faint emission. The flux map was integrated over a 200\,km\,s$^{-1}$ window. As shown in the bottom left panel of Figure~\ref{fig:images}, [C\,\textsc{ii}] emission is detected from ZD3+ZD6 (unresolved) and ZD12. Their total fluxes were similarly measured with \texttt{imfit}. 

The measured [C\,\textsc{ii}] fluxes were converted to [C\,\textsc{ii}] luminosities, which were then used to estimate the molecular gas mass, $M_{\rm gas}$. Following previous studies \citep[e.g.,][]{2024A&A...682A..24A,2025arXiv250110508A}, we adopted the empirical relation from \citet{2018MNRAS.481.1976Z}, which links [C\,\textsc{ii}] luminosity to molecular gas mass based on a diverse galaxy sample at $z \sim 0.5$--6:
\begin{equation}
\log \left( \frac{L_{\mathrm{[CII]}}}{L_\odot} \right) = -(1.28 \pm 0.21) + (0.98 \pm 0.02) \log \left( \frac{M_{\mathrm{H}_2}}{M_\odot} \right)
\end{equation}
Atomic gas is not taken into account in our analysis, as in \citet{2025arXiv250110508A}. Derived quantities for the sources are summarized in Table~\ref{tab:measurements}. The estimated molecular gas masses span ${\rm log}(M_{\rm gas}/M_\odot) \sim 9.0$--9.6, approximately 1\,dex lower than those of the REBELS sample at $z \sim 7$ \citep{2024A&A...682A..24A,2025arXiv250110508A}.

The 1.26\,mm continuum image traces rest-frame $\sim$142\,$\mu$m emission at $z \approx 7.88$. As shown in the top-right panel of Figure~\ref{fig:images}, dust continuum is detected toward YD1, YD4+YD6, and YD7W. The peak S/N measured in the tapered map is 4.4, 5.1, and 3.4, respectively. We measured fluxes using \texttt{imfit} for YD1 and YD4+YD6, while YD7W is marginally detected in the tapered map, and its flux was estimated from the peak pixel value. These detections confirm the earlier report, which had lower significance \citep{2023ApJ...955L...2H}.
In the {\it Chain} region, no dust continuum emission is identified (Figure~\ref{fig:images}, top-left panel). We therefore place $3\sigma$ upper limits on the fluxes of ZD3+ZD6 and ZD12 using the tapered map, assuming unresolved sources.

Dust masses ($M_{\rm dust}$) were estimated from the rest-frame 142\,$\mu$m photometry assuming an optically thin modified blackbody. Following \citet{2025arXiv250110508A}, we adopt a fixed dust temperature of $T_{\rm dust} = 45 \pm 15$\,K\footnote{A possible dust-continuum detection at 362\,GHz was reported by \citet{2017ApJ...837L..21L}. From our re-analysis, we do not confirm the detection and instead place a $3\sigma$ point-source limit of $190\,\mu{\rm Jy}$ (synthesized beam $0\farcs66\times0\farcs51$). This limit does not provide a meaningful constraint on $T_{\rm dust}$, although $T_{\rm dust}=45\pm15$\,K remains allowed.}
. We also assume a dust emissivity index $\beta = 2.0$ and an opacity coefficient $\kappa_0 = 10.41$\,g$^{-1}$\,cm$^{2}$ at $\nu_0 = 1900$\,GHz, consistent with previous studies \citep{2022MNRAS.512...58F,2023MNRAS.518.6142A,2024MNRAS.533.3098A,2025arXiv250110508A}. Corrections for cosmic microwave background (CMB) effects at $z = 7.88$ were applied following \citet{2013ApJ...766...13D}.
The derived values are listed in Table~\ref{tab:measurements}. The dust masses span ${\rm log}(M_{\rm dust}/M_\odot) \sim 6.0$--6.4, approximately 1\,dex lower than those of the REBELS sample and A1689-zD1 at $z \sim 7$ \citep{2015Natur.519..327W,2021MNRAS.508L..58B,2024A&A...682A..24A,2025arXiv250110508A}.

We also compile stellar mass \citep{2025arXiv250706284W} and gas-phase metallicity measurements from the literature \citep{2023ApJ...947L..24M,2025ApJ...985...83M,2024A&A...691A..19V}, as summarized in Table~\ref{tab:measurements}. 
Stellar masses are derived from the multi-band photometry using \texttt{Prospector}, adopting a nonparametric star-formation history (SFH) \citep{2025MNRAS.537..112W,2025arXiv250706284W}.
We adopt gas-phase metallicities derived using strong-line methods involving [O\,\textsc{ii}], H$\beta$ and [O\,\textsc{iii}], such as the $R_{23}$ index and \(\hat{R}\) \citep{2024A&A...681A..70L}. We note that YD7W lacks an H$\beta$ detection and is therefore excluded from the metallicity-related diagram.

\begin{figure}
  \centering
  \includegraphics[width=1.0\columnwidth]{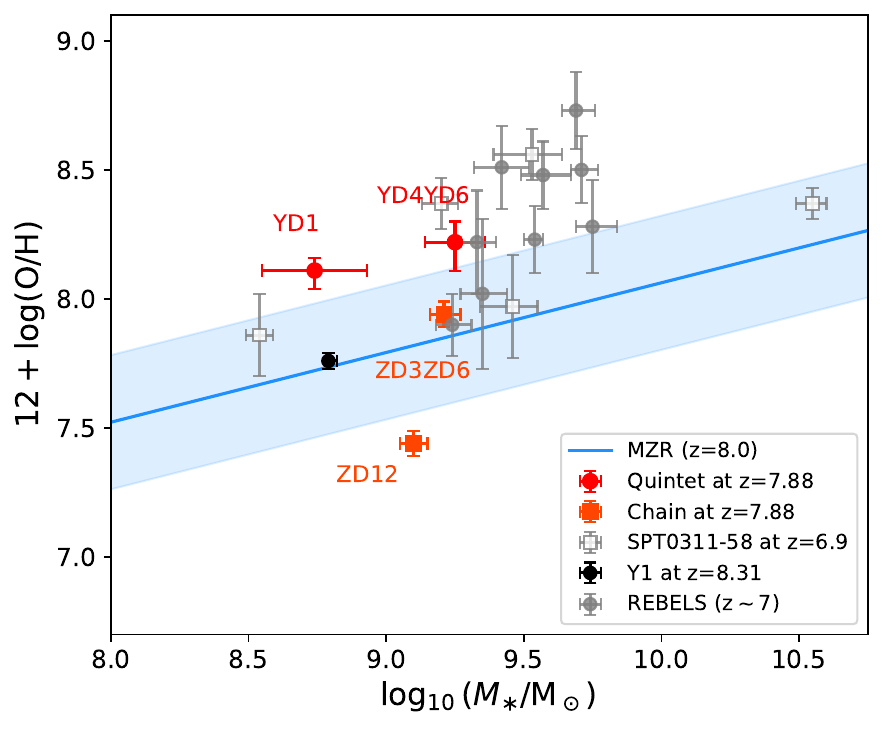}
\caption{
Distribution of galaxies in the A2744-z7p9OD proto-cluster and other $z \sim 7$--8 samples \citep{2024ApJ...975...87M, 2024ApJ...977L..36H, 2024A&A...688A.146A, 2025arXiv250110508A, 2025arXiv250110559R} in the stellar mass--metallicity plane. All metallicities are derived using strong-line calibrations. The mass--metallicity relation from \citet{2024ApJ...971...43M} is shown for comparison as a blue shaded region. YD1 and YD4$+$YD6 in the {\it Quintet} exhibit elevated gas-phase metallicities relative to the relation, while ZD3$+$ZD6 and ZD12 in the {\it Chain} lie on or below the relation.
}
\label{fig:mzr}
\end{figure}

\section{Discussion} \label{sec:dis}

\begin{figure}
  \centering
  \includegraphics[width=1.0\columnwidth]{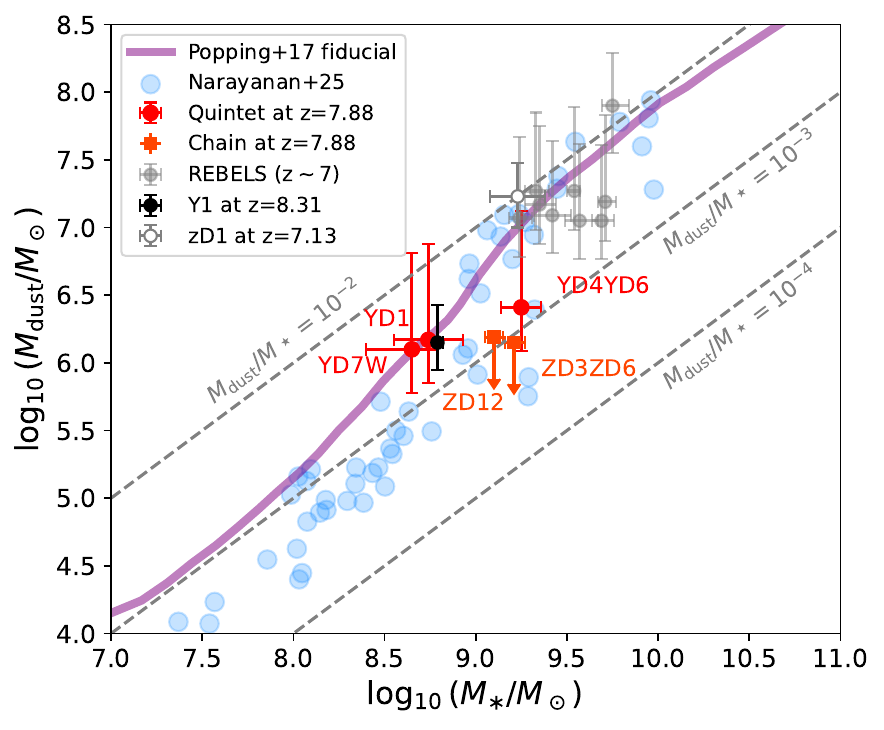}
  \caption{Comparison of dust mass ($M_{\rm dust}$) versus stellar mass ($M_\star$) for our sample and other $z \sim 7$--8 galaxies \citep{2015Natur.519..327W, 2020MNRAS.493.4294B,2021MNRAS.508L..58B,2024ApJ...975...87M,2025arXiv250110508A}, along with model predictions at $z=7$ \citep{2017MNRAS.471.3152P} and at $z=6-12$ \citep{2025ApJ...982....7N}. The A2744-z7p9OD proto-cluster galaxies occupy the lower stellar mass regime and show relatively low dust-to-stellar mass ratios, $\log (M_{\rm dust}/M_\star)$, compared to the REBELS sample at $z\sim7$. Simulations predict that a rapid rise in this ratio occurs around $\log (M_\star/\mathrm{M}_\odot) \sim 9$ at $z \sim 8$, driven by efficient dust growth via metal accretion in the ISM. The galaxies in the {\it Quintet} and {\it Chain} may represent this transitional phase.}
  \label{fig:msmd}
\end{figure}

\begin{figure}
  \centering
  \includegraphics[width=1.0\columnwidth]{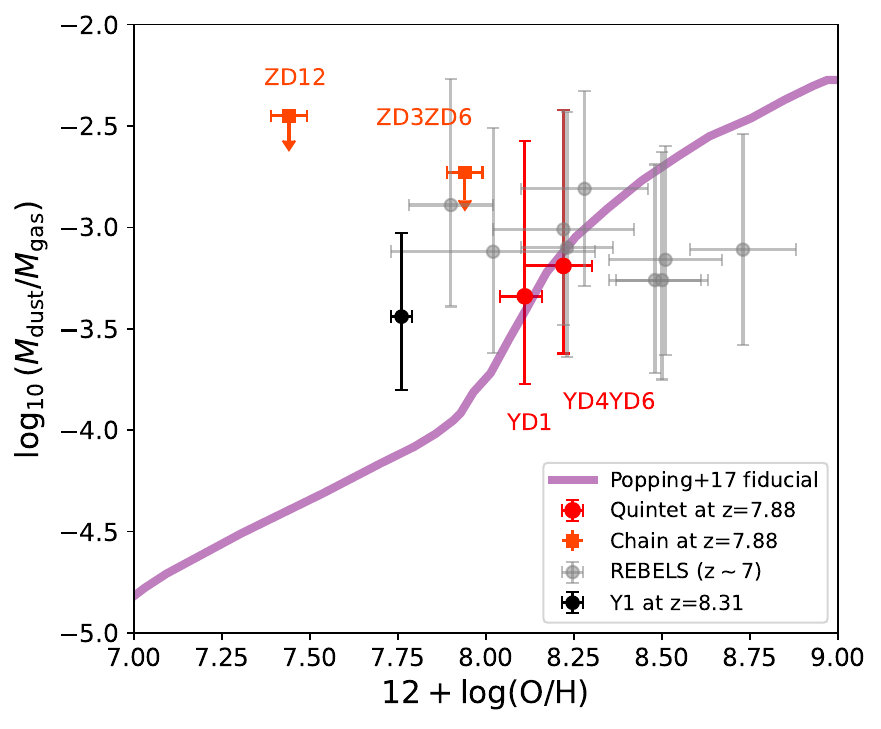}
\caption{
Dust-to-gas mass ratio, $\log (M_{\rm dust}/M_{\rm gas})$, as a function of gas-phase metallicity for galaxies in our sample and comparison samples \citep{2024ApJ...975...87M, 2024ApJ...977L..36H, 2025arXiv250110508A}. The solid curve indicates the predicted relation from (semi-)analytical calculations of \citet{2017MNRAS.471.3152P} at $z=7$. YD1 and YD4+YD6 in the Quintet exhibit relatively high metallicities, comparable to those of the REBELS sample, suggesting that they may be undergoing rapid dust mass buildup driven by grain growth in the ISM.
}
\label{fig:gdr}
\end{figure}

Figure~\ref{fig:mzr} summarizes the stellar-mass--gas-phase metallicity relation (``mass-metallicity relation'') for the {\it Quintet} and {\it Chain} groups in the A2744-z7p9OD proto-cluster. For comparison, we show the mass--metallicity relation derived for galaxies at $3<z<9.5$ from \citet{2024ApJ...971...43M}, along with data for the REBELS galaxies  $z\sim7$. 
We also include MACS0416-Y1 at $z = 8.31$ (hereafter Y1; \citealt{2019ApJ...874...27T,2020MNRAS.493.4294B,2023ApJ...952....9T,2024ApJ...975...87M,2024ApJ...977L..36H}) as an additional $z \approx 8$ reference in the discussion below, completing the current sample of four most distant galaxies with dust continuum detections.

Interestingly, ZD3+ZD6 and ZD12 in the {\it Chain} exhibit metallicities consistent with or slightly below the relation at the redshift, whereas YD1 and YD4+YD6 in the {\it Quintet} lie above it, although these galaxies have similar stellar masses. This unveils a diversity in metal enrichment across group environments within the proto-cluster core (see also \citealt{2024ApJ...971...43M}), with the {\it Quintet} showing the most advanced chemical evolution. Such a diversity in metallicity is also reported in the SPT0311-58 overdensity at $z=6.9$ \citep{2024A&A...688A.146A}.
While systematic uncertainties from strong-line calibrations introduce non-negligible scatter, all metallicities were derived using a similar method involving [O\,\textsc{ii}], H$\beta$ and [O\,\textsc{iii}]. 
%
%
Therefore, uncertainties due to methodological differences are minimal, and the relative trends are likely robust. Among the comparison samples, Y1 lies on the relation, while the REBELS galaxies fall on or above it, suggesting that the REBELS galaxies are relatively metal-enriched systems at slightly lower redshifts than the $z\sim8$ targets.

Figure~\ref{fig:msmd} shows the relation between stellar mass and dust mass. In addition to the $z \sim 7$--8 measurements, predictions from theoretical works \citep{2017MNRAS.471.3152P,2025ApJ...982....7N} are also shown. Both account for dust production from SNe and AGB stars, although the contribution from AGB stars is expected to be subdominant at $z \gtrsim 6$ \citep[e.g.,][]{2015MNRAS.451L..70M}, as well as for dust growth via metal accretion in the ISM.
\citet{2025ApJ...982....7N} trace the evolution of the $M_{\rm dust}$--$M_\star$ relation from $z \sim 12$ to $z \sim6$, reproducing the observed dust-to-stellar mass ratios of $\log (M_{\rm dust}/M_\star) \sim -2$ at $z \sim 6$--7. Their model also predicts significantly lower values at $z \gtrsim 10$ ($\log (M_{\rm dust}/M_\star) \sim -3$ to $-4$), followed by a rapid rise in the dust content around $z \sim 8$ at $\log (M_\star/M_\odot) \sim 9$, driven by efficient dust growth through metal accretion in the ISM. A similar trend is seen in the predictions of \citet{2017MNRAS.471.3152P} and other studies \citep[e.g.,][]{2022MNRAS.512..989D, 2023MNRAS.519.4632D, 2023MNRAS.526.4801T,2023MNRAS.525.4976M}.

Interestingly, the $z\sim8$ galaxies ({\it Quintet}, {\it Chain}, and Y1) fall within the predicted transition regime between metal-poor systems at $z\gtrsim10$ and the more dust-rich galaxies at $z \sim7$ like the REBELS samples. A substantial scatter is also observed: the {\it Chain} galaxies appear less dusty than those in the {\it Quintet}, consistent with the diversity in the transition stage expected in the models.  This trend persists even when adopting alternative mass--metallicity relations from the literature (e.g., \citealt{2024A&A...684A..75C}).
Following \citet{2015A&A...577A..80M} , we estimate the per-SN dust yield (see Appendix \ref{app:yield}). For the \textit{Quintet} galaxies, the required yield is $\simeq 0.1$--$0.2~M_\odot$ per SN, which would be consistent with a dust model by a supernova origin only if most of the newly formed dust survives the reverse shock. However, in reality, a significant fraction of SN-driven dust is expected to be destroyed, implying that this yield likely requires additional grain growth in the ISM to account for the observed dust masses as suggested by \citet{2015A&A...577A..80M}. For the \textit{Chain} galaxies, the upper limits ($<0.1~M_\odot$ per SN) are compatible with an SN origin.
These results suggest that we may be witnessing the predicted transition phase in dust mass assembly—from early production by SNe to efficient dust growth via metal accretion in the ISM.

By combining ALMA and JWST measurements, we are now able to construct a comprehensive dataset covering gas-phase metallicity, dust mass, and molecular gas mass. Figure~\ref{fig:gdr} presents the relation between gas-phase metallicity and the dust-to-gas mass ratio (DGR; $\log (M_{\rm dust}/M_{\rm gas})$) for galaxies in the A2744-z7p9OD proto-cluster, along with other $z = 7$--8 samples and theoretical predictions.
As shown earlier, YD1 and YD4+YD6 in the {\it Quintet} are more metal-rich than ZD3+ZD6 and ZD12 in the {\it Chain}, which appears to correlate with the detectability of their dust continuum. The {\it Quintet} galaxies exhibit gas-phase metallicities comparable to those of the REBELS galaxies. Such chemically evolved ISM conditions (i.e., higher metal content) may promote efficient dust growth through metal accretion, increasing the fraction of solid dust among total ISMs, as predicted by simulations \citep[e.g.,][]{2017MNRAS.471.3152P}.

We note several caveats that currently limit the interpretation of our results. One of the most significant is the uncertainty in dust temperature. Assuming $T_{\rm dust} = 30$--60\,K introduces an uncertainty of up to $\sim$1\,dex in $M_{\rm dust}$, which may partially obscure underlying trends. If $T_{\rm dust} \gg 60$\,K, as suggested for Y1 \citep{2020MNRAS.493.4294B}, the inferred dust masses would be even lower than estimated here, implying that we may be observing an earlier phase of dust enrichment.
The current sensitivity may also be insufficient to detect dust continuum in low-metallicity systems. Deeper, multi-band follow-up observations would be beneficial to further validate this scenario and refine current constraints. 
Additionally, converting [C\,\textsc{ii}] luminosity to molecular gas mass introduces further systematic uncertainty. We assess its impact by re-deriving \(M_{\rm mol}\) using two recent simulation-based calibrations (Appendix~\ref{app:gasmass}; \citealt{2025A&A...693A.119C,2025arXiv250414001V}). These prescriptions include dependences on redshift and gas-phase metallicity, and can drive \(M_{\rm mol}\) either upward or downward relative to our fiducial choice. Across our sample, the resulting \(M_{\rm mol}\) values differ by \(\sim 0.3\text{--}0.5\) dex, yet the overall trend in Fig.~\ref{fig:gdr} remains unchanged.

Lastly, we consider the role of the environment in metal and dust enrichment. Dust detection at $z\sim8$ is still rare. Within the Abell2744-z7p9OD proto-cluster, dust is detected in the Quintet, whereas no dust is found in the Chain. Recent studies based on spectral energy distribution (SED) fitting suggest that several galaxies in the {\it Quintet} exhibit signs of quenching and past episodes of active star formation 
\citep[][see also Y.Fudamoto et al. submitted, W. Osone et al. in prep.]{2025arXiv250706284W}, whereas such systems are much rarer in the {\it Chain}.
Although the {\it Quintet} and the {\it Chain} together appear to constitute the core of the proto-cluster (Figure~\ref{fig:images}), the {\it Quintet} likely marks the true core, having experienced a longer and more intense star-formation history. This may have led to more advanced chemical enrichment in the ISM and CGM. These results suggest that dense proto-cluster environments can play a crucial role in early dust production, as also proposed by \citet{2023ApJ...955L...2H}. The deeper gravitational potential enhances gas accretion from the cosmic web, fueling accelerated star formation and enrichment cycles.
Indeed, at $z \sim 3$, dusty star-forming galaxies (DSFGs) have been observed to cluster in the dense cores of proto-clusters, aligned along cosmic web filaments \citep{2015ApJ...815L...8U,2019Sci...366...97U}. Conversely, galaxies in less dense environments, such as voids, may retain a more pristine ISM compared to their counterparts in the general field. Such a trend can emerge at $z\gtrsim7-8$ (e.g., \citealt{2024A&A...688A.146A}).

\begin{deluxetable*}{ccccccccc}
\tablecaption{Summary of [C\,\textsc{ii}], continuum, and derived quantities\label{tab:summary}}
\tablehead{
\colhead{Group} &
\colhead{ID} &
\colhead{$S_{\mathrm{[CII]}}\Delta v$} &
\colhead{$\log(M_{\mathrm{mol}}/M_\odot)$} &
\colhead{$S_{1.26\,\mathrm{mm}}$} &
\colhead{$\log(M_{\mathrm{dust}}/M_\odot)$} &
\colhead{$\log(M_\star/M_\odot)^{1}$} &
\colhead{$12 + \log \mathrm{(O/H)}^{2,3}$} &
\colhead{$\mu^{1}$}
}
\startdata
Quintet & YD1      & $154\pm14$ & $9.51\pm0.29$ & $15.4\pm3.5$ & $6.17_{-0.32}^{+0.71}$ & $8.74_{-0.19}^{+0.19}$ & $8.11_{-0.07}^{+0.05}$ & $1.94$ \\
Quintet & YD4+YD6  & $190\pm19$ & $9.60\pm0.29$ & $26.7\pm4.4$ & $6.41_{-0.32}^{+0.71}$ & $9.25_{-0.11}^{+0.11}$ & $8.22_{-0.11}^{+0.08}$ & $1.95$ \\
Quintet & YD7W     & $140\pm14$ & $9.47\pm0.29$ & $13.2\pm3.8$ & $6.10_{-0.32}^{+0.71}$ & $8.65_{-0.25}^{+0.14}$ & \ldots                 & $1.96$ \\
Chain   & ZD3+ZD6  & $69\pm10$  & $9.17\pm0.29$ & $<14.2$      & $<6.15$                 & $9.21_{-0.05}^{+0.06}$ & $7.94_{-0.05}^{+0.05}$ & $1.87$ \\
Chain   & ZD12     & $44\pm19$  & $8.98\pm0.34$ & $<15.4$      & $<6.19$                 & $9.10_{-0.05}^{+0.05}$ & $7.44_{-0.05}^{+0.05}$ & $1.86$ \\
\enddata
\tablecomments{[C\,\textsc{ii}] and continuum fluxes are given in Jy\,km\,s$^{-1}$ and $\mu$Jy, respectively, and are not corrected for gravitational lensing. Molecular gas and dust masses are lensing-corrected using the magnification factors listed in the table. References: (1) \cite{2025arXiv250706284W}, (2) Venturi et al. (2024), (3) \cite{2025ApJ...985...83M}.}
\label{tab:measurements}
\end{deluxetable*}

\begin{acknowledgments}
We thank the anonymous referee for constructive comments and suggestions.
We acknowledge fruitful discussions with Raffaella Schneider.
This work is based on observations made with the NASA/ESA/CSA James Webb Space Telescope. The data were obtained from the Mikulski Archive for Space Telescopes at the Space Telescope Science Institute, which is operated by the Association of Universities for Research in Astronomy, Inc., under NASA contract NAS 5-03127 for JWST. These observations are associated with program \#2561.
This paper makes use of the following ALMA data: ADS/JAO.ALMA\#2018.1.01332.S, \#2023.1.00193.S,\#2023.1.01362.S. ALMA is a partnership of ESO (representing its member states), NSF (USA) and NINS (Japan), together with NRC (Canada), NSTC and ASIAA (Taiwan), and KASI (Republic of Korea), in cooperation with the Republic of Chile. The Joint ALMA Observatory is operated by ESO, AUI/NRAO and NAOJ.
HU acknowledges support from JSPS KAKENHI Grant Numbers 25K01039. This work was supported by NAOJ ALMA Scientific Research Grant Numbers 2024-26A.
YN acknowledge funding from JSPS KAKENHI Grant Number 23KJ0728. L.Colina acknowledges support by grant PIB2021-127718NB-100 from the
Spanish Ministry of Science and Innovation/State Agency of Research
MCIN/AEI/10.13039/501100011033 and by “ERDF A way of making Europe” 
MH is supported by Japan Society for the Promotion of Science (JSPS) KAKENHI Grant No. 22H04939.

\end{acknowledgments}

\appendix
\restartappendixnumbering

\section{Additional Notes on Molecular Gas Mass}
\label{app:gasmass}

The conversion between the observed [C\,\textsc{ii}] line luminosity and the molecular gas mass is still under debate, and different prescriptions have been proposed in the literature. To reflect this uncertainty, in addition to the values based on \citet{2018MNRAS.481.1976Z} listed in Table~\ref{tab:measurements}, we also report molecular gas masses derived from two recent simulation-based calibrations \citep{2025A&A...693A.119C,2025arXiv250414001V} in Table~\ref{tab:mgas2}.

\citet{2025A&A...693A.119C} propose the following relation between [C\,\textsc{ii}] luminosity and molecular gas mass:
\begin{equation}
\label{eq:cii_mh2}
\log_{10}\!\left(\frac{L_{[\mathrm{C\,II}]}}{L_\odot}\right)
= 1.36(\pm 0.01)\,\log_{10}\!\left(\frac{M_{\mathrm{H}_2}}{M_\odot}\right)
- 0.14(\pm 0.01)\, z
- 4.13(\pm 0.09)\, .
\end{equation}

\citet{2025arXiv250414001V} instead express the conversion as
\( M_{\mathrm{H}_2} = \alpha_{[\mathrm{C\,II}]}\, L_{[\mathrm{C\,II}]} \),
with a metallicity-dependent factor
\begin{equation}
\label{eq:alpha_cii_unc}
\log_{10}\!\left(\frac{\alpha_{[\mathrm{C\,II}]}}{M_\odot\,L_\odot^{-1}}\right)
= \left(-0.39 \pm 0.06\right)\,
\log_{10}\!\left(\frac{Z}{Z_\odot}\right)
+ \left(0.67 \pm 0.06\right)\, .
\end{equation}

\begin{deluxetable}{ccc}
\tablecaption{Summary of molecular-gas mass measurements\label{tab:mgas_summary}}
\tablehead{
\colhead{ID} &
\colhead{$\log(M_{\mathrm{mol,\,C25}}/M_\odot)$} &
\colhead{$\log(M_{\mathrm{mol,\,V25}}/M_\odot)$}
}
\startdata
YD1     & $9.76\pm0.12$ & $8.94\pm0.10$ \\
YD4+YD6 & $9.83\pm0.12$ & $8.99\pm0.12$ \\
YD7W    & $9.73\pm0.12$ & \ldots        \\
ZD3+ZD6 & $9.58\pm0.12$ & $8.67\pm0.11$ \\
ZD12    & $9.38\pm0.18$ & $8.68\pm0.22$ \\
\enddata
\tablecomments{$M_{\mathrm{mol,\,C25}}$ and $M_{\mathrm{mol,\,V25}}$ denote the [C\,\textsc{ii}]-based molecular gas masses derived using the calibrations of Casavecchia et al.\ (2025) and \citet{2025arXiv250414001V}, respectively.}
\label{tab:mgas2}
\end{deluxetable}

\section{[C\,\textsc{ii}] flux measurement on the \textit{Quintet}}
\label{app:imfit}

\begin{figure*}
  \centering
  \includegraphics[width=1.0\columnwidth]{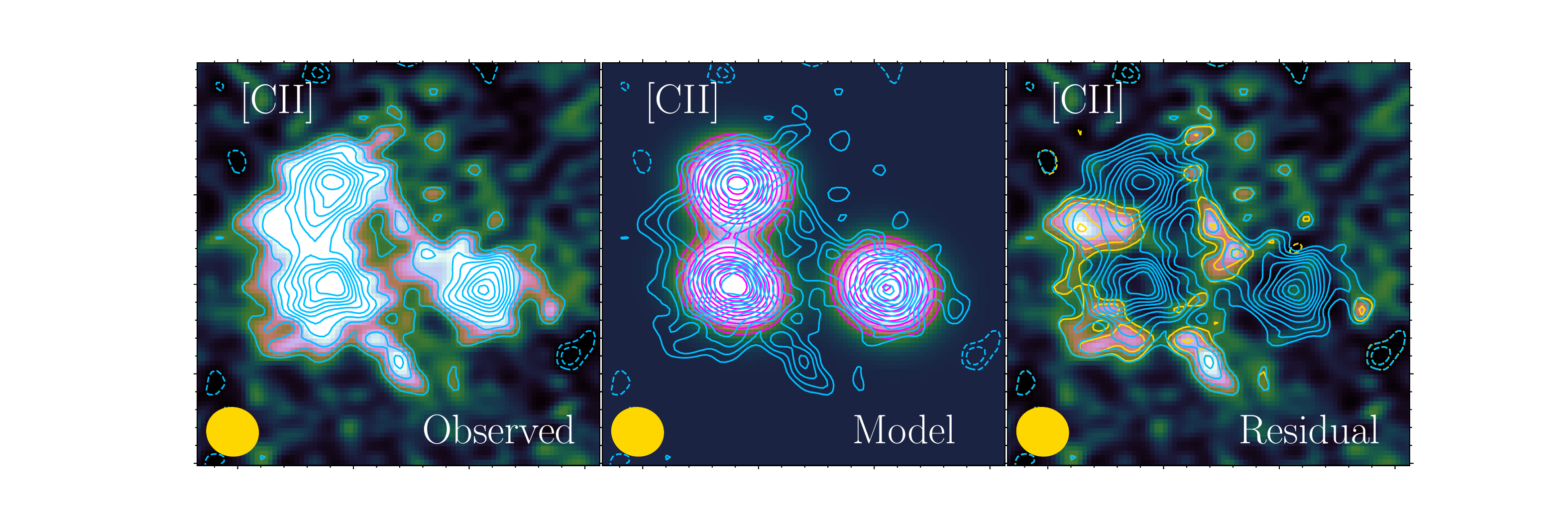}
\caption{
Results of the multi-component Gaussian fit to the Quintet map. The left, middle, and right panels show the observed image, the best-fit model, and the residual (data$-$model), respectively. Three components (YD1, YD4$+$YD6, and YD7W) are modeled. Cyan, magenta, and yellow contours are drawn for observed, modeled, and residual images, respectively, at $\pm2\sigma$, $\pm3\sigma$, $\ldots$.
}
\label{fig:imfit}
\end{figure*}

We add notes on how the fluxes of the \textit{Quintet} members (YD1, YD4+YD6, and YD7W) are measured. 
We first create a flux-integrated image by collapsing the cube over a velocity range of 
$\Delta v = 360\,\mathrm{km\,s^{-1}}$. 
We then run \texttt{imfit} with position-specific masks to fit the emission directly associated with each galaxy. 
The results are shown in Fig.~\ref{fig:imfit}. 
The fits perform well overall, although some extended emission remains in the residual map. 
For YD4+YD6, the residual emission overlaps the stellar counterpart (part of YD6), and we add this contribution to the total flux of YD4+YD6. 
Further discussion of the extended emission bridging the galaxies is presented in Fudamoto et al., submitted.

\section{Dust yield}
\label{app:yield}

We compute the per--supernova (SN) dust yield following \citet{2015A&A...577A..80M}:
\begin{equation}
\label{eq:yield}
Y_{\rm dust}^{\rm SN} \;=\; f \times \left(\frac{M_{\rm dust}}{M_{\star}}\right),
\end{equation}
where $f$ is an IMF-dependent factor. For a Chabrier IMF \citep{2003ApJ...586L.133C} in the SN case, $f=84$.
The resulting yields (in $M_\odot$ per SN) are
$0.23$, $0.12$, $0.24$, $<0.07$, and $<0.10$ for YD1, YD4$+$YD6, YD7W, ZD3$+$ZD6, and ZD12, respectively.

\bibliography{reference_2024a}{}
\bibliographystyle{aasjournal}



\end{document}